%% file: GBP-ISIT2025-v3.tex
\documentclass[conference]{IEEEtran}
\usepackage{epsfig}
\usepackage{times}
\usepackage{float}
\usepackage{afterpage}
\usepackage{amsmath}
\usepackage{amstext}
\usepackage{amssymb,bm}
\usepackage{latexsym}
\usepackage{color}
\usepackage{array,algpseudocode}
\usepackage{makecell}
\usepackage{diagbox}
\usepackage{amsmath}
\usepackage{amsthm}
\usepackage{graphicx}

\usepackage[center]{caption}
\usepackage{pstricks}
\usepackage{subfigure}
\usepackage{booktabs}
\usepackage{enumerate}
\usepackage[ruled]{algorithm}
\usepackage{romannum}
\usepackage{derivative}
\usepackage{cite}
\usepackage{lipsum}
\usepackage{cuted}
\usepackage{stackengine}

\usepackage{bbm}
\def\BibTeX{{\rm B\kern-.05em{\sc i\kern-.025em b}\kern-.08em
    T\kern-.1667em\lower.7ex\hbox{E}\kern-.125emX}}
\usepackage[normalem]{ulem}
\usepackage{url}

\newtheorem{lemma}{Lemma}
\newtheorem{definition}{Definition}

\newtheorem{proposition}{Proposition}

\newcommand{\uS}{\underline {S}} 
\newcommand{\us}{\underline {s}} 
\newcommand{\uY}{\underline {Y}} 
\newcommand{\uy}{\underline {y}}
\newcommand{\uX}{\underline {X}} 
\newcommand{\Qsf}{\mathsf{Q}} 
\newcommand{\ux}{\underline {x}}

\newcommand{\comment}[1]{}
\newcommand{\floor}[1]{\lfloor #1 \rfloor}

\newcommand{\var}[1]{\text{var}{#1}}

\newcommand\blfootnote[1]{
    \begingroup
    \renewcommand\thefootnote{}\footnote{#1}
    \addtocounter{footnote}{-1}
    \endgroup
}

\let\svthefootnote\thefootnote
\newcommand\blankfootnote[1]{%
  \let\thefootnote\relax\footnotetext{#1}%
  \let\thefootnote\svthefootnote%
}

\input{macros.tex}

\allowdisplaybreaks

\begin{document}
\title{ \huge{ On the Sensing Capacity  of Gaussian ``Beam-Pointing'' Channels with Block Memory and Feedback} }
\input{authors.tex}
\maketitle

\begin{abstract}
Driven by the demands of high-frequency wireless communications in 5G and 6G systems (e.g., mmWave, sub-THz), we explore a state-dependent  {\em Gaussian beam-pointing} (GBP) channel. In this model, the channel state defines an unknown angle of departure (AoD), which remains constant within each coherence block of  $\Qsf$ time slots but changes independently across blocks.
The transmitter receives strictly causal feedback which may originate from a radar detection system or explicit feedback from the receiver at the end of each slot and estimates the AoD at the end of each block.
To enhance transmission efficiency, we propose a joint communication and sensing scheme. While the communication capacity of the GBP channel has been previously analyzed by the authors, this work focuses on sensing capacity, characterized by the mutual information between the channel state and the feedback conditioned on the transmitted signal.   
We derive an upper bound using dynamic programming and propose an achievable inner bound on the sensing capacity, both formulated as optimization problems. For the special case of $\Qsf = 1$, the proposed transmission scheme achieves the optimal sensing rate and highlights the inherent trade-off between sensing and communication performance.

\end{abstract}

\section{Introduction}
\label{sec:intro}
The rapid growth of 5G and the anticipated deployment of 6G wireless networks have intensified research into millimeter-wave (mmWave) and sub-terahertz (sub-THz) frequencies to support ultra-high data rates   \cite{Rappaport2019Above100G,Ghosh20195GEvolution,uusitalo20216g}. 
However, these frequency bands suffer from severe path loss, making reliable communication viable primarily in line-of-sight (LoS) scenarios, necessitating the deployment of large antenna arrays and advanced beamforming techniques to focus signal energy toward the receiver \cite{Rappaport2017}. 
\blfootnote{The work of Siyao Li was partially supported by The Boeing Center for Aviation and Aerospace Safety at Embry-Riddle Aeronautical University and by the European Research Council under the ERC Advanced Grant No. 789190 (CARENET). The work of Shuangyang Li was supported in part by the European Union’s Horizon 2020 Research and Innovation Program under MSCA Grant No. 101105732 (DDComRad). The work of Giuseppe Caire was supported by BMBF Germany under the program of ``Souverän. Digital. Vernetzt.'' Joint Project 6G-RIC (Project ID 16KISK030).}  

The Gaussian Beam-Pointing (GBP) channel \cite{ISIT2024}, which extends the binary beam-pointing channel previously studied by the authors \cite{BBP:TIT2023}, provides a tractable yet insightful framework for analyzing the information-theoretical performance of beamforming-based communication systems. In this model, the transmitter, using a large antenna array, aims to detect and communicate with a receiver at an unknown angle of departure (AoD) \cite{ISIT2024}. The AoD, treated as a discrete channel state, remains constant over  $\Qsf$  time slots within each coherence block but varies independently between blocks.\footnote{In practical mobile-radio channels, the multipath angles and delays experience a much slower rate of variation than the fast-fading amplitudes, and can be considered stationary over several data transmission blocks.}  Unlike traditional beam acquisition (BA) schemes that separate beam alignment and data transmission \cite{alkhateeb2017initial,Sequential,Javidi-AL,Low-complex}, this work investigates a joint communication and sensing (JCAS) approach to improve overall efficiency, particularly under a limited number of time slots within each coherence block.

  Our focus is on maximizing sensing performance from an information-theoretic perspective. Prior research has analyzed the communication capacity of GBP channels assuming large signal dimensions per time slot \cite{ISIT2024}.
 The feedback capacity of finite-state Markov channels was investigated in \cite{finite-state-machine,Capacity-Achieving-Liu2015} from a closed-loop control perspective. The authors of \cite{Opportunistic2010} worked on the error exponent and capacity of each individual time slot of a compound channel, and presented reasonable error exponents using training sequence. 
The fundamental trade-off between communication rate and sensing Cramér-Rao bound (CRB) in Gaussian channels without feedback was revealed in \cite{Fundamental-Gaussian,LiuFanTradeoff-ISIT}.   Related works evaluating communication and sensing performance with dedicated receivers are presented in \cite{UnifiedFramework,Joudeh2022Joint,CovertJCAS,SecureISAC} and references therein. Studies in \cite{Zhang2011Joint,kobayashi2018joint,ahmadipour2021informationtheoretic} focus on memoryless channels, while \cite{Capacity-Estimation-2005} examines the JCAS problem over a state-dependent channel with states known at the transmitter. More relevant to this work, \cite{Rate-Distortion2023} investigates the asymptotic state detection error exponent of the mono-static model with an unknown but fixed state during the whole transmission.  In contrast, our dynamic scenario where states remain the same only for finite time slots,  characterized by in-block memory (iBM) \cite{Kramer-memory}, requiring  performance optimization over the coherence interval.

\paragraph*{Contribution}  
To address the challenges of fundamental sensing-communication trade-offs in beamforming under practical resource constraints (e.g., power and time), we study the GBP channel with feedback and block memory under a per-slot power constraint. We formulate the sensing capacity as a dynamic programming problem and demonstrate that the posterior probability of the channel states guides the determination of the optimal input distribution at each time slot. 
Building upon the JCAS strategy, we tackle the challenge of optimal source design, focusing on beam selection for transmitting Gaussian codewords based on the posterior state probabilities.  We derive the corresponding achievable sensing rates and examine through numerical examples, the inherent trade-off between communication and sensing capacities, as well as the efficiency of our proposed scheme. 

\paragraph*{Notations} 
For an integer $n$, we let $[n] = \{ 1, \cdots, n\}$ and $[n_1: n_2] = \{ n_1, \cdots, n_2\}$ for some integers $n_1 < n_2$.  $\underline{X}$ denotes a vector and $\underline{X}^n = [\underline{X}_1, \cdots, \underline{X}_n ]$ denotes a sequence of vectors. $\| \underline X\|$ denotes the $l_2$ norm of $\underline X$. 
Let   $\underline{1}_m$ denote the one-hot vector with the single one appearing at index $m$, and $\mathbbm{1}{\{ \cdot \}}$ denotes an indicator function. We use $\bar{x}$ to denote the logical negation of the binary variable $x$. 
The complex circularly symmetric Gaussian distribution is represented by $\mathcal {CN}(\mu, \sigma^2)$, where the probability density function for $\mu \in \mathbb{C}$ and $x \in \mathbb{C}$ is given as $\frac{1}{\pi \sigma^2} e^{-\frac{\| x - \mu\|^2}{\sigma^2}}$.
$\floor{ x}^+$ takes the greatest positive integer less than or equal to $x$.  ${\mathbb I}_{q}$ is a $q \times q$ identity matrix.  Let $H( \cdot )$ denote the binary entropy function. 

 \section{System Model and Definitions}
\label{sec:generalized}
For the GBP channel model, 
the channel state $\uS \in \{0,1\}^M$ is a one-hot vector with the ``1'' element placed with uniform probability over the $[M]$ positions, indicating the beam direction  ``containing" the receiver.  
Each sample is referred to as signal dimension, or channel use, of the underlying waveform channel. \footnote{In real-world wireless communications time is slotted and packets of $q$ signal dimensions (complex samples) are transmitted as modulated waveforms. } 
Let $n = \ell \Qsf$ denote the total transmission time in channel symbols, 
where each symbol has a dimension of $q$, and $\ell$ is the number of state coherence blocks spanned by one message transmission.
A strictly causal feedback $Z$ is provided to the transmitter for each forward channel symbol. 
The channel statistic is characterized by 
$P_{ \uS^\ell,  \uY^n,  Z^n, \uX^n }( \us^\ell, \uy^n, z^n, \ux^n) \!\!
= \prod_{i=1}^\ell \!\!P_{\uS}( \us_i \!)  P_{ \underline X_i^{\Qsf}\|Z_i^{\Qsf}}( \underline x_i^{\Qsf} \| z_i^{\Qsf}\!)  \!
 P_{ \uY_{i,j}, Z_{i,j} | \uX_{i,j},  \uS_i } (  \uy_{i,j},z_{i,j}| \ux_{i,j},  \us_i ), $ 
where (using the notation of \cite{Kramer-memory}) 
\begin{equation}
\!\!\!\! P_{ \underline X_i^{\Qsf}\|Z_i^{\Qsf}}( \underline x_i^{\Qsf} \| z_i^{\Qsf})\! =\! \prod_{j = 1}^{\Qsf} \!P_{  \underline X_{i,j} |  \underline X_{i,1}^{j-1}, Z_{i,1}^{j-1} }(  \underline x_{i,j} |  \underline x_{i,1}^{j-1}, z_{i,1}^{j-1}), \label{directed-P}
\end{equation}
and the received signal at channel use $j$  in $i$-th block is 
\begin{align}
\underline Y_{i,j} = {\underline S}_i^T {\underline X}_{i,j} + \underline N_{i,j}, \label{eq:channel-model}
\end{align} 
where $\underline X_{i,j} \in \mathbb C^{M \times q}$ is the 
space-time signal matrix transmitted at slot $j$ of block $i$, 
the channel state ${\underline S}_i$  known by the receiver remains the same for the whole block and follows a uniform distribution with $P_{\uS_i}(\underline 1_m) = \frac{1}{M}$, $m \in [M]$, 
and $\underline N_{i,j} \sim \mathcal{CN}({\bf 0}, {\mathbb I}_{q\times q})$ is the complex additive white Gaussian noise (AWGN) for all $i \in [\ell]$, $j \in [\Qsf]$. 
The binary feedback signal $Z_{i,j} \in \{0, 1\}$ is obtained at the end of each slot $j \in [\Qsf]$ as the result of a threshold test on the received signal energy, given by 
\begin{align}
Z_{i,j} \triangleq \mathbbm{1}{\{  \| \underline Y_{i,j} \|^2 > \nu  \}}, 
\label{eq:feedback}
\end{align} 
and serves as a one-bit indicator to inform the transmitter if the direction of the receiver 
is detected or not with some error probability that depends on the feedback threshold $\nu$ and the signal-to-noise ratio (SNR).  We wish to emphasize that the feedback function (\ref{eq:feedback}) is part of the channel definition, and  the threshold $\nu$ is also fixed \textit{a priori}.  
 Given the past channel inputs $\underline X_i^{j-1}$ and feedback $Z_i^{j-1}$, the current channel input is statistically 
independent of the channel state $\underline S_i$, i.e., $\underline S_i \mk (\underline X_i^{j-1}, Z_i^{j-1}) \mk \underline X_{i,j}$. 

\subsection{Definitions}
Based on the available encoding functions at the transmitter, we can determine the channel input $ \underline X_i^{\Qsf}$ 
based on the message $W$ and the causal feedback $Z_i^{\Qsf-1}$.   
Moreover, the estimation is not only causal, but also restricted on a per-block basis.\footnote{ Since the channel state is independent and identically distributed (i.i.d.) across coherence blocks, it is natural that the optimal estimator for the channel described in \eqref{directed-P}-\eqref{eq:feedback}, under a time-averaged distortion criterion, would decompose into separate estimators for each block. In this work, we explicitly impose the per-block state estimation as a system constraint, which can be justified by the fact that sensing must be provided in a timely 
fashion to accomplish some desired real-time system function (e.g., positioning, or similar).} 
This prompts the following definition of state estimation:
\begin{definition}\normalfont
\label{def:stateEstimation}
Let the estimated state sequence $\hat{  \underline S}^\ell$ from the (known) message $W$ and (causally but not strictly causally 
known) feedback $Z^{\ell \Qsf}$ is given by 
 $\hat{ \underline S}_{i} \triangleq e_i(  W, Z_i^{\Qsf}), \ i \in[\ell], $ 
for sequence of state estimation functions $e_i(\cdot): \mathcal W \times \mathcal Z^{\Qsf} \to \hat{\mathcal S}$
where $\hat{\mathcal S}$ is the reproduction alphabet.
\hfill $\lozenge$
\end{definition}
Based on the definition given above, in this work we use the  {\em sensing capacity} to quantity the sensing performance: 
\begin{definition}\normalfont
\label{def:sensing-rate}
The sensing rate $\Upsilon(B, \Qsf)$ of the GBP with iBM, input cost $B$ and state coherence block length $\Qsf$ with associated per-block state estimation is given by  
\begin{equation} 
\Upsilon(B, \Qsf) = \frac{1}{\ell} \sum_{i=1}^{\ell}I( \underline S_i; Z_i^{\Qsf} | \underline X_i^{\Qsf}),
\end{equation} 
 and the sensing capacity is given by
\begin{equation} 
\Omega(B,\Qsf)  = \max_{P_{ \underline X_i^{\Qsf}\|Z_i^{\Qsf}} \in \Bc}  \; \frac{1}{\ell} \sum_{i=1}^{\ell} I( \underline S_i; Z_i^{\Qsf} |  \underline X_i^{\Qsf}) \label{sensing-rate}
\end{equation}
where 
$P_{\underline S_i, Z_i^{\Qsf},  \underline X_i^{\Qsf}} = P_{ \underline S_i} P_{ \underline X_i^{\Qsf}\|Z_i^{\Qsf}} \prod_{j=1}^\Qsf P_{Z_{i,j} |  \underline S_i, \underline X_{i,j}  } $
and $\Bc$ is the set of distributions in the form (\ref{directed-P}) satisfying 
$\mathbb{E}[ b(  \underline X_{i,j}) ] \leq B$ for all $i \in [\ell], j \in [\Qsf]$. \hfill $\lozenge$
\end{definition} 
Note that the sensing capacity in Definition \ref{def:sensing-rate} does not assume any channel coding. Hence, we shall refer to 
$\Omega(B,\Qsf)$ as the {\em sensing-only} capacity, symmetrically with 
the communication (only) capacity. Since $\underline S \sim P_{\uS}$ is independent of $\underline X^{\Qsf}$, implying that   
$I( \underline S; Z^{\Qsf} | \underline X^{\Qsf}) = H(\underline S) - H(\underline S|Z^{\Qsf}, \underline X^{\Qsf})$, the sensing rate is uniquely determined by the 
conditional entropy $H(\underline S|Z^{\Qsf},\underline X^{\Qsf})$. This is shown in \cite{HamdiG2024} 
to {\em precisely} quantify the optimal
performance in terms of log-loss of a state {\em soft estimator} in a JCAS context, 
producing a posterior distribution of $ \underline S$ given $(Z^{\Qsf},  \underline X^{\Qsf})$. 
Hence, maximizing $I( \underline S; Z^{\Qsf} |  \underline X^{\Qsf})$ (sensing rate) corresponds to minimizing the log-loss of state soft estimation.
Based on the i.i.d. channel state, in the rest of the paper, we eliminate the block index $i$ and consider 
\begin{equation} 
\Upsilon(B, \Qsf) = \max_{P_{ \underline X^{\Qsf}\|Z^{\Qsf}} \in \Bc}  I( \underline S; Z^{\Qsf} |  \underline X^{\Qsf}) . \label{eq:sensing-original}
\end{equation}

Furthermore, we define the true alignment indicator
 \begin{align}
 T_{j}  = f(\uS, \underline X_{j}) \triangleq \mathbbm{1}_{\{ \frac{1}{q}\| \uS^T \underline X_{j} \| > 0\}},  \label{eq:T}
 \end{align}
representing whether a non-zero row of the transmitted signal matrix reaches the receiver, $j \in [\Qsf]$. The true indicator sequence $T^j  = f^{(j)}(\uS, \underline X^j)$ takes $T_{k}$ provided by \eqref{eq:T} for all $k\in[j]$.  It is possible that the transmitted signal aligns with the beam direction, i.e., $T=1$, while the noisy feedback indicates not aligned $Z=0$ and vice versa. 
Recall that the receiver determines the binary feedback $Z$ based on the power of the received signal. Accordingly, we define two performance metrics for one-bit feedback as follows.
\begin{definition}\normalfont
\label{def:FA-MD}
Based on the observation of channel output $\underline Y$, let $\mathcal H_0$ represent the event that $\underline Y = \underline N$, i.e.,  noise only, and $\mathcal H_1$ represent the event that $\underline Y = \underline S^T \underline X+ \underline N$, i.e.,  signal-plus-noise.
The false alarm (FA) and miss detection (MD) probabilities are respectively defined as
$P_\text{FA} =  P[Z_{j} =1 | \mathcal H_0 , \uS = \us],$ 
$ P_\text{MD} = P[Z_{j} =0 |\mathcal H_1, \uS = \us ].$ 
 \hfill $\lozenge$
\end{definition}
As $I(\underline S; Z^{\Qsf} |  \underline X^{\Qsf} ) \leq I( \underline S; T^{\Qsf} |  \underline X^{\Qsf})$,
smaller FA and MD probabilities leads to larger sensing capacity. Moreover, 
\begin{align}
P_\text{FA} = P( \| \underline Y\|^2 > \nu | \mathcal H_0, \us)= 1 - F_{\mathcal X_{2q}^2}(\nu) 
\label{eq:PFA}
\end{align} where $\mathcal H_0:  \underline Y \sim \prod_{i=1}^q f_0(y_i)$, and  $F_0 \sim \mathcal{CN}(0, 1)$. 
 We assume that the system parameter $\nu$ is fixed and chosen such that the false alarm probability satisfies $P_{\text{FA}} < 0.5$ throughout the transmission.
Moreover, under some power constraint allocated to the CSI (e.g., $\sigma^2_{m}$, $m\in [M]$), by \eqref{eq:feedback}, we have the minimum $P_\text{MD}$ as follows when the channel input is known at the receiver
\begin{align}
P_\text{MD} = P( \| \underline Y\|^2 \leq \nu | \mathcal H_1, \us) = F_{\mathcal X_{2q}^2(\sigma_m^2)}(\nu) 
\label{eq:PMD}
\end{align}
which is the CDF of a chi-square distribution with dimension $2q$ and non-centralized parameter $\sigma_m^2$ at $\nu$. \footnote{If the channel input is pilot signal, it is trivial to have \eqref{eq:PMD}. If each row of $\underline X_{j}$ is Gaussian and the sensor is unaware of the transmitted codeword, the receiver uses only its empirical statistics, then the hypothesis test reduces to the energy threshold test (e.g., see  \cite{Poor1994book}). }  Therefore, given signal dimension $q$ and power threshold $\nu$, $P_\text{FA}$ is a constant and $P_\text{MD}$ is a function of the power assigned on the AoD.

\section{Main Results}
\label{sec:main}
In this section, we formulate the sensing capacity in \eqref{eq:sensing-original} as a dynamic programming problem, propose a JCAS transmission scheme, and derive the corresponding achievable sensing rate based on the effective state variable defined in the dynamic programming formulation. 


\subsection{Sensing Capacity Outer Bound}
\label{subsec:outer}
As discussed before, we just need to focus on 
\begin{align}
&\min_{P_{ \underline X^{\Qsf}\|Z^{\Qsf}} \in \Bc}  H(\underline S| \underline X^\Qsf, Z^\Qsf  ) \notag
 %
 \\& =   \sum_{\underline x^{\Qsf}} \sum_{ z^{\Qsf} \us} P_{\uS}(\us) P_{\underline X^\Qsf \| Z^{\Qsf}}(\underline x^\Qsf \| z^{\Qsf}) \prod_{j=1}^\Qsf P_{ Z_j| \underline X_j, \uS}(z_j | \underline x_j, \us) \notag
 \\& \cdot \log  \frac{ \sum_{\underline s^{\prime}} \prod_{j=1}^{\Qsf} P_{Z_j | \underline S, \underline X_j}(z_i | \underline s^{\prime}, \underline x_j )}{\prod_{j=1}^{\Qsf} P_{ Z_j| \underline X_j, \uS}(z_j | \underline x_j, \us)  },  \label{eq:mutual-info}
    \\&
\text{ where } P_{\uS | \underline X^{\Qsf}, Z^{\Qsf} }(\us |  \underline x^{\Qsf}, z^{\Qsf}) 
 \notag 
\\& = \frac{ \prod_{j=1}^{\Qsf} P_{Z_j | \underline S, \underline X_j}(z_j | \underline s, \underline x_j ) P_{\underline X^{\Qsf} \|  Z^{\Qsf} }(\underline x^{\Qsf} \| z^{\Qsf}) P_{\uS}(\us) }{ \sum_{\underline s^{\prime}} \prod_{j=1}^{\Qsf} P_{Z_j | \underline S, \underline X_j }(z_j | \underline s^{\prime}, \underline x_j ) P_{\underline X^{\Qsf} \|  Z^{\Qsf} }(\underline x^{\Qsf} \| z^{\Qsf}) P_{\uS}(\us^{\prime}) } \notag 
 \\& =  \frac{\prod_{j=1}^{\Qsf} P_{ Z_j| \uS, \underline X_j }(z_j | \us, \underline x_j  )  }{ \sum_{\underline s^{\prime}} \prod_{j=1}^{\Qsf} P_{Z_j | \underline S, \underline X_j  }(z_j | \underline s^{\prime}, \underline x_j  )},
 \label{eq:state-simplified}
        \end{align}
        and $P_{ Z_j| \underline X_j, \uS}(z_j | \underline x_j, \us)$  is a function of the power allocated to the CSI by \eqref{eq:PFA} and \eqref{eq:PMD}.  Given the design of probing signals, \eqref{eq:mutual-info}  implies that the state detection is fundamentally limited by how distinguishable the states are through the channel input and binary feedback, conditioned on the signal energy allocation.
      
         \begin{definition} \normalfont
          \label{def:Ki}
          Let $\underline K_i$ be the random vector of the posterior probabilities of the channel state $\uS$ at time $i$, whose realization $\underline k_i$ is 
         $ \underline k_i = [ k_i(1), k_i(2), \cdots, k_i(M)]$ 
          where $ k_0(m) = \frac{1}{M}$ and 
     $k_i(m) = P_{\uS | \underline X^{i}, Z^{i} }(\us = \underline 1_m | \underline x^{i}, z^{i}),  \  \forall m \in [M].$  \hfill  $\lozenge$
        
          \end{definition}
          
 Note that $ k_i(m)$ can be represented by Bayes' rule as in  \eqref{eq:psp-recursive}.   
 \begin{table*}     
   \begin{align} 
k_i(m) 
 &=\frac{ k_{i-1}(m) P_{ Z_i | \underline X_i, \uS }(z_i | \underline x_i, \underline 1_m)   P_{ \underline X_i | \underline X^{i-1}, Z^{i-1} }(\underline x_i | \underline x^{i-1}, z^{i-1} ) }{ \sum_{m \in [M]} k_{i-1}(m) P_{ Z_i | \underline X_i, \uS }(z_i | \underline x_i, \underline 1_m )   P_{ \underline X_i | \underline X^{i-1}, Z^{i-1} }(\underline x_i | \underline x^{i-1}, z^{i-1} )} 
 = \frac{ k_{i-1}(m) P_{ Z_i | \underline X_i, \uS }(z_i | \underline x_i, \underline 1_m)   }{ \sum_{m \in [M]}  k_{i-1}(m) P_{ Z_i | \underline X_i, \uS }(z_i | \underline x_i, \underline 1_m )  }.
 \label{eq:psp-recursive}
    \end{align}
\end{table*}
    We refer to the posterior distribution $P_{\uS | \underline X^{i}, Z^{i} }(\us | \underline x^{i}, z^{i})$ as the \textit{state} used by the transmitter to generate the  channel input distribution $P_{ \underline X_i | \underline X^{i-1}, Z^{i-1} }(\underline x_i | \underline x^{i-1}, z^{i-1} )$, referred to as the \textit{policy}. 
    
    \begin{lemma} 
    \label{lemma:input-dis-replace}
    The  posterior distribution $\underline k_{t-1}$  can be used to replace $(\underline x^{t-1}, z^{t-1})$ for the purpose of determining the optimal  input distribution  $P_{ \underline X_t | \underline X^{t-1}, Z^{t-1} }(\underline x_t | \underline x^{t-1}, z^{t-1} )$ at time $t$, i.e.,
     \begin{align*}
  P_{ \underline X_t | \underline X^{t-1}, Z^{t-1} }(\underline x_t | \underline x^{t-1}, z^{t-1} ) =   P_{ \underline X_t | \underline K_{t-1} }(\underline x_t | \underline k_{t-1} ).
    \end{align*}
        \end{lemma}
    \begin{IEEEproof}
   The proof follows similar steps as outlined in \cite{finite-state-machine} and is omitted here for brevity. 
              \end{IEEEproof}
 We now formulate the sensing capacity calculation as a finite-horizon stochastic control problem, where dynamic programming is employed to determine the optimal sequence of transmission policies that maximize the expected sensing rate over the coherence block.  
 At each time step $t$, we use the posterior belief $k_{t-1}$ as the system state and computes the candidate policy $\pi_t(k_{t-1}) \triangleq P_{ \underline X_t | \underline K_{t-1} }(\underline x_t | \underline k_{t-1} ) $ that maximizes the mutual information by Lemma   \ref{lemma:input-dis-replace}. 
  Denote   $\underline \pi = [\pi_1, \cdots, \pi_{\Qsf}]$.  
   The system equation of this dynamic system is \eqref{eq:psp-recursive}, and we represent it as 
        $ \underline k_t = \text{Func}(\underline k_{t-1}, P_{\underline X_{t}| \underline K_{t-1}}(\underline X_t | \underline k_{t-1}), z_t)$
           where the system feedback $Z_t$ follows conditional distribution as given in Definition \ref{def:FA-MD}. 
       Let the \textit{reward} for each stage $t$ be
       \begin{align*}
       & r_t(\underline k_{t-1},  P_{\underline X_t | \underline K_{t-1}}( \underline X_t | \underline k_{t-1}), Z_t) = I(\uS; \underline X_{t}, Z_{t} | \underline k_{t-1} ) 
       \\&  \hspace{4.7cm} = I(\uS; \underline X_{t}, Z_{t} | \underline x^{t-1}, z^{t-1} ) 
      \\
& \text{     where }    P_{\uS, \underline X_{t}, Z_{t} | \underline K_{t-1}}(\us, \underline x_{t}, z_{t} | \underline k_{t-1} ) & 
     \\ & \hspace{1cm} = P_{\uS, \underline X_{t}, Z_{t} | \underline X^{t-1}, Z^{t-1}}(\us, \underline x_{t}, z_{t} | \underline x^{t-1}, z^{t-1}) &
       \\  & \hspace{1cm} = P_{\uS | \underline X^{t-1}, Z^{t-1}}(\us | \underline x^{t-1}, z^{t-1}) &
       \\&  \hspace{1cm} \cdot P_{\underline X_{t} | \underline X^{t-1}, Z^{t-1}} (\underline x_{t} |  \underline x^{t-1}, z^{t-1}) P_{Z_t | \underline X_t, \uS} (z_t | \underline x_t, \us) &
       \\&  \hspace{1cm} = \underline k_{t-1} P_{ \underline X_t | \underline K_{t-1}}(\underline x_t | \underline k_{t-1} ) P_{Z_t | \underline X_t, \uS} (z_t | \underline x_t, \us). &
       \end{align*}
       As $ I(\uS; Z^{\Qsf} | \underline X^{\Qsf}) = \sum_{i=1}^{\Qsf} I(\uS; \underline X_{i}, Z_{i} | \underline X^{i-1}, Z^{i-1})$, the expectation of the average reward per stage equals to the sensing rate $ I(\uS; Z^{\Qsf} | \underline X^{\Qsf})$. We denote 
        \vspace{-0.3cm}
        \begin{align*}
      J(\underline \pi, P_{\uS}) &= \mathbb{E}_{\underline \pi}[ \sum_{t=1}^{\Qsf} I(\uS; \underline X_{t}, Z_{t} | \underline x^{t-1}, z^{t-1} ) | P_{\uS} ]
      \\& =  \mathbb{E}_{\underline \pi}[ \sum_{t=1}^{\Qsf} r_t(\underline k_{t-1}, P_{\underline X_t | \underline K_{t-1}}( \underline X_t | \underline k_{t-1}), Z_t) | \underline k_0]
             \end{align*}
              \vspace{-0.05cm}
\!\!where $\underline \pi$ is the control policy used, $k_0$ is a given initial state,      and the goal is to maximize $J(\underline \pi, P_{\uS})$ over all control policies, and find an optimal policy $\underline \pi^{\star}$ that achieves the maximal reward $J^\star( P_{\uS})$. Therefore, 
   $ J^\star( P_{\uS})= $ $ J(\underline \pi^\star, P_{\uS})$ $=\!\!\! \max_{\underline \pi \in \mathcal B} J( \underline \pi, P_{\uS})$
          where $\mathcal B$ is the set of all input distribution satisfying the short-term power constraint presented in Definition \ref{def:sensing-rate}. 
       Finding the optimal source distribution that maximizes the sensing rate is an average-reward-per-stage stochastic control problem \cite{DP}. 
       Define the  reward to go function
        $V( \underline k_{t-1}) =  \mathbb{E}[\sum_{i=t}^{\Qsf} r_t( P_{\underline X_t | \underline K_{t-1}}( \underline X_t | \underline k_{t-1}), Z_t) | \underline K_{t-1} = \underline k_{t-1} ]. $ 
        Observe that $V( \underline k_0) = J(\underline \pi, P_{\uS})$, and express $V( \underline k_{t-1})$ 
        as a Bellman's equation
          $ V(\underline K_{t-1} ) = \max_{  \pi_t  \in \mathcal B }  
         \{  \mathbb{E}[r_t( \underline K_{t-1}, \pi_t, Z_t )]  
    + \mathbb{E}[ V(  \text{Func}(\underline K_{t-1}, \pi_t, Z_t) ) ] \}.$
 A simple dynamic programming value-iteration algorithm \cite{DP} is presented in Algorithm \ref{alg:dynamic-programming} to find the input distribution that maximizes the sensing rate $I(\uS; Z^{\Qsf} | \underline X^{\Qsf})$ through posterior tracking.
 
  \begin{algorithm}[t] 
\caption{Optimal Input Distribution to Maximize Sensing Rate}
\label{alg:dynamic-programming}
\begin{algorithmic}[1]
 \State {\bf Initialization:} 
At the terminal stage $t=\Qsf$, compute the value function by maximizing over the terminal policy $ \pi_{\Qsf}$ such that 
          $ V(\underline K_{\Qsf-1} ) = \max_{  \pi_{\Qsf}  \in \mathcal B }  
        \mathbb{E}[ V(  \text{Func}(\underline K_{\Qsf-1}, \pi_{\Qsf}, Z_{\Qsf}) ) ]$
       where  
         $  \pi_{\Qsf}^{\star} = \text{arg} \max_{  \pi_{\Qsf}  \in \mathcal B }  
        \mathbb{E}[ V(  \text{Func}(\underline K_{\Qsf-1}, \pi_{\Qsf}, Z_{\Qsf}) ) ]. $
\State {\bf Recursions:}  
  \For{$ t=   1:\Qsf$} 
  	\State Compute the $t$-stage reward-to-go function as 
           $V(\underline K_{t-1} ) = \max_{  \pi_{t}  \in \mathcal B }  
         \{ r_t( \underline K_{t-1},  \pi_t , Z_t )  + \mathbb{E}[ V(  \text{Func}(\underline K_{t-1},  \pi_t , Z_t) ) ] \}. $
	\State The optimized policy is taken as
          $\pi_{t}^{\star} = \max_{  \pi_{t}  \in \mathcal B }  
         \{ r_t( \underline K_{t-1}, \pi_t , Z_t ) 
     + \mathbb{E}[ V(  \text{Func}(\underline K_{t-1},  \pi_t , Z_t) ) ] \}.$
\EndFor
\State {\bf Output:}  Starting from $t=1$, use the optimal policy sequence $\underline \pi^{\star} = \{ \pi_1^{\star}, \cdots,  \pi_{\Qsf}^{\star} \}$ to compute the total expected reward. 
\end{algorithmic}
\end{algorithm}

In general, for given signal dimension $q$, power threshold $\nu$, memory block length $\Qsf$, and power constraint $B$, the sensing capacity does not admit a closed form. However, for the memoryless case, i.e., $\Qsf = 1$, 
   we have a memoryless channel, obtain \eqref{eq:-H-condition-Q=1}, 
    \begin{table*}
          \begin{align}
          H(\underline S| \underline X_1, Z_1  ) \notag
         & = \frac{1}{M}\sum_{\underline x_1} P_{\underline X_1}(\underline x_1)  \sum_{\us} \left(  P_{Z_1| \underline X_1, \uS}(1|\underline x_1, \us) 
          \log  \frac{ \sum_{\underline s^{\prime}}  P_{Z_1 | \underline S, \underline X_1 }(1 | \underline s^{\prime}, \underline x_1 )}{ P_{ Z_1| \underline X_1, \uS}(1 | \underline x_1, \us)  } 
           + (1- P_{Z_1| \underline X_1, \uS}(1 |\underline x_1, \us) )
          \log  \frac{ \sum_{\underline s^{\prime}}  (1-P_{Z_1 | \underline S, \underline X_1 }(1 | \underline s^{\prime}, \underline x_1 ) )}{ 1- P_{ Z_1| \underline X_1, \uS}(1 | \underline x_1, \us)  } \right) \notag
          \\& = \frac{1}{M}\sum_{\underline x_1} P_{\underline X_1}(\underline x_1)   \Big( M \log M +  \sum_{\us}H(P_{ Z_1| \underline X_1, \uS}(1 | \underline x_1, \us) ) 
          - M H(  \frac{ \sum_{\us } P_{ Z_1| \underline X_1, \uS}(1 | \underline x_1, \us)}{M}) \Big).
          \label{eq:-H-condition-Q=1}
                    \end{align}
                    \vspace{-0.7cm}
                     \end{table*} 
                    and the sensing rate becomes
\begin{align}
&H(\uS) - H(\uS | \underline X_1, Z_1) = \sum_{\underline x_1} P_{\underline X_1}(\underline x_1)  \notag
\\& \cdot \Big( H(  \frac{ \sum_{\us } P_{ Z_1| \underline X_1, \uS}(1 | \underline x_1, \us)}{M}) - \frac{\sum_{\us}H(P_{ Z_1| \underline X_1, \uS}(1 | \underline x_1, \us) )}{M}    \Big) \label{eq:Rs-Q=1}
\\& \leq H( P_{Z_1}(1)) \!\!- \!\! \sum_{\underline x_1} P_{\underline X_1}(\underline x_1)\frac{\sum_{\us}H(P_{ Z_1| \underline X_1, \uS}(1 | \underline x_1, \us) )}{M}  \label{eq:jensen-ineq}
\end{align}  
where  
\eqref{eq:Rs-Q=1}  is independent of the indices of the beams selected but depends on the number of beams and the power allocated to the selected beam, and \eqref{eq:jensen-ineq} follows from Jensen's inequality. Therefore, the sensing rate is independent of the choice of the transmitted beam but depends on the number of beams and the power allocated to them.  
Intuitively, there is a tradeoff between the number of beams to probe and the amount of power allocated to the selected beams. 

\subsection{Sensing Capacity Inner Bound}
\label{subsec:inner}
As discussed in \cite{ISIT2024}, the communication capacity is achieved by uniformly distributing power across all possible beams when $\Qsf =1$. However, this approach does not hold for sensing capacity. Instead, the transmitter must select a subset of the beams and allocate power strategically. 
 The proposed beam selection policy is guided by the posterior distribution $P_{\uS|\underline X_j, Z_j}$, which is recursively updated based on Bayes’ rule \eqref{eq:psp-recursive}. This selection rule is consistent with the principle of maximizing information gain, as it focuses probing energy on the most likely state hypotheses.
Additionally, we observe that the MD probability remains unchanged,  as indicated in \eqref{eq:PMD}, when the transmitted signal on the selected beam follows a Gaussian distribution with zero mean and variance $\frac{\sigma_m^2}{q} \mathbb{I}_q$.  
Therefore, rather than transmitting deterministic signals, we use information-carrying signals for both sensing and communication, aligning perfectly with the JCAS philosophy. 
In the following, we introduce a coded posterior beam acquisition scheme, similar to Algorithm 1 proposed in \cite{ISIT2024}, with the key difference being the beam selection process after each feedback. Specifically, we select beam indices that provide the highest posterior state probabilities, as outlined in Lemma~\ref{lemma:input-dis-replace}. 

For the first transmission, the transmitter randomly and uniformly selects $n_{z^0}$ beams and assigns power $q\frac{B}{n_{z^0}}$ to each selected beam, where the initial number of beams to be probed denoted by $n_{z^0}$ is selected such that both the FA and MD probabilities are greater than 0.5. The selected beams are stored in set $\mathcal B_{j}^e, \forall j \in [\Qsf]$. Recall that by \eqref{eq:PFA},  $\nu > 2q$ ensures that $P_\text{FA} > 0.5$. Then, given $q$, we just need to determine $\sigma_m^2$ to obtain the MD probability as shown in \eqref{eq:PMD}.  
At the end of each transmission, the posterior probability $P_{\uS | \underline X^j, Z^j}$ is updated by \eqref{eq:psp-recursive} and $d_{z^j}$   beams with highest posterior probability will be stored in set $\mathcal B_{z^j}(\underline x^{j})$, and initially $d_{z^0} = M$. For the $(j+1)$-th transmission ($j\geq 1$), the transmitter randomly and uniformly selects $n_{z^{j}}$ beams from these $d_{z^j}$ beams, and sends complex Gaussian signals with zero mean and variance $b_{z^j} \mathbb I_q$ where $b_{z^j} \geq \frac{B}{n_{z^0}}$.  
Inspired by the sensing capacity achievable scheme of the binary beam-pointing channel under the per-slot power constraint \cite{BBP:TIT2023}, we consider a specific choice of the parameters  $n_{z^{j-1}}$  as follows 
\begin{align}
 n_{z^{j-1} } &= 
\begin{cases}
0  &  d_{z^j} = 0
\\
 \max( \min(\floor{\frac{d_{z^j }}{2}}^{+}, n_{z^0}), 1) & \text{o.w.}
\end{cases}. \
\label{eq:semi-bi-sec-d-update}
\end{align}

 
 Furthermore, all the possible beam indices providing true indicator sequence $t^j$ based on the transmitted signal $\underline x^j$ will be stored in set  $\mathcal B_{t^j}(\underline x^j)$, and $\mathcal B_{t^j}(\underline x^j)$ is updated as follows
\begin{align}
\mathcal B_{t^j}(\underline x^j) = \begin{cases}
\mathcal B_{t^{j-1}}(\underline x^{j-1}) \cap \mathcal B_j^e, & t_j =1
\\
\mathcal B_{t^{j-1}}(\underline x^{j-1}) \backslash \mathcal B_j^e, & t_j =0
\end{cases}. \label{eq:update-m-t}
\end{align} 
Note that the sets $\mathcal B_{t^j }(\underline x^j) $ are exclusive for $t^j$ and $\sum_{t^j} | \mathcal B_{t^j}(\underline x^j) | = M$.  If we add a super script $t^j$ to $\mathcal B_{z^j}(\underline x^{j})$ (i.e., $\mathcal B_{z^j}^{t^j}(\underline x^{j})$) to represent the sequence of $t^j = f^{(j)}( \us, \underline x^j) $ for  $\us \in \mathcal B_{z^j}(\underline x^{j})$, we have $\mathcal B_{z^j}^{t^j}(\underline x^{j}) = \mathcal B_{z^j}(\underline x^{j}) \cap \mathcal B_{t^j}(\underline x^{j})$, and $|\mathcal B_{z^j}(\underline x^{j})| = \sum_{t^j} |\mathcal B_{z^j}^{t^j}(\underline x^{j})|$, that is $d_{z^j} = \sum_{t^j} d_{z^j}^{t^j}$.

 According to the selection of the number of beams, we have  $P_{Z_1 | \uS, \underline X_1}(z_1 = t_1 | \us, \underline x_1, t_1) > 0.5 > P_{Z_1 | \uS, \underline X_1}(z_1= \bar{t_1} | \us, \underline x_1, t_1)$, which leads to $ \mathcal B_{z^1}(\underline x^1) =   \mathcal B_{t^1}(\underline x^1)$ where $t^1 = z^1$. For the following transmission, the power assigned to the selected beams will be lower bounded by $b_{z^0}$, which ensures $P_{Z_j | \uS, \underline X_j}(z_j = t_j | \us, \underline x_j, t_j) > 0.5 > P_{Z_j | \uS, \underline X_j}(z_j = \bar{t_j} | \us, \underline x_j, t_j)$ holds for all $j\in [\Qsf]$. Hence,  by \eqref{eq:psp-recursive} $ \mathcal B_{z^j}(\underline x^j) =   \mathcal B_{t^j}(\underline x^j)$  where $t^j = z^j$, and $\mathcal B_{z^j}^{t^j}(\underline x^j) = \emptyset$ where $t^j \neq z^j$. 
Moreover, one can easily verify that now  the number of beams stored in set $ \mathcal B_{t^j}(\underline x^j)$  is independent of $\underline x^j$ given $t^j$, which provides
\begin{align}
\!\!\!\!\! P_{Z_j | T_j}( z_j | t_j) &= \begin{cases}
1 - F_{\mathcal X_{2q}^2}(\nu) & \text{ if } z_j = 1, t_j = 0 
\\
F_{\mathcal X_{2q}^2(b_{z^{j-1}} )}(\nu)  & \text{ if } z_j = 0, t_j = 1 
\end{cases},
\label{eq:condition-Z-T}
\end{align}
and $P_{Z_j | T_j}( z_j = t_j | t_j) = 1 - P_{Z_j | T_j}( z_j  = \bar{t_j} | t_j) $.

Denoting $m_{t^j} = |\mathcal B_{t^j}(\underline x^j)|$ and initializing $m_{t^0} = M$, we obtain that $m_{t^j} = (2 t_j-1) n_{z^{j-1}=t^{j-1}} + (1-t_j) m_{t^{j-1}}$ by \eqref{eq:update-m-t} and also, $d_{z^j} = \sum_{t^j} d_{z^j}^{t^j} = d_{z^j}^{t^j=z^j} = m_{t^j=z^j}$. Thus,  $d_{z^j}$ is updated as $d_{z^j} = (2z_j-1) n_{z^{j-1}} + (1-z_j) d_{z^{j-1}}$. 
To satisfy the short-term power constraint, we need the following holds 
\begin{align}
& \sum_{z^{j-1}, t^{j-1} } n_{z^{j-1}} b_{z^{j-1}} \frac{m_{t^{j-1}}}{M}  \prod_{k=1}^{j-1} P_{ Z_k | T_k }( z_k | t_k )  
 \leq qB \label{eq:power-constraint}  
\end{align} 
for all $j\in[\Qsf]$, where $P_{Z_k| T_k}(z_k| t_k)$ is given by \eqref{eq:condition-Z-T} and $\prod_{k=1}^{0}P_{Z_k| T_k}(z_k| t_k) =1$.  Continuing with \eqref{eq:mutual-info}, the sensing rate can be simplified as follows. 

\begin{proposition}
\label{lemma:bi-section}
The  sensing rate $R_{\rm s}$ of the posterior coded beam acquisition scheme 
with sequences of $\{ n_{z^{j}}\}$ given in \eqref{eq:semi-bi-sec-d-update} is the solution to the optimization problem 
 \begin{align*}
\max_{n_{z^0}, b_{z^j}, \forall j \in [\Qsf]}& \sum_{z^{\Qsf}} \sum_{t^{\Qsf}} \frac{m_{t^{\Qsf}}}{M} \prod_{j=1}^{\Qsf}P_{Z_j| T_j}(z_j| t_j) 
 \notag
\\& \cdot \log\left( \frac{ M \prod_{j=1}^{\Qsf}P_{Z_j | T_j }(z_j | t_j) }{\sum_{t^\Qsf} m_{t^{\Qsf}} \prod_{j=1}^{\Qsf}P_{Z_j | T_j}(z_j | t_j) }  \right)
\\
\text{s.t. }  &\eqref{eq:power-constraint},   
  \forall j \in[\Qsf]. 
\end{align*}
\end{proposition}
 \paragraph*{Proof Sketch of Proposition 1}
The optimization problem in Proposition 1 is derived by explicitly writing the mutual information $I(\uS; Z^{\Qsf} | \underline X^{\Qsf})$ under the proposed policy, where each beam index update and feedback response is modeled through binary decisions influenced by the predetermined $\nu$. The form of the objective directly follows from this mutual information computation using the chain rule, and the constraints ensure compliance with the short-term power constraint.


While the paper formally defines and analyzes the sensing capacity, the practical relevance lies in understanding how resources (e.g., power, time, beamspace) are shared between sensing and communication tasks. In particular, in the memoryless case ($\Qsf = 1$), our results clearly demonstrate a fundamental trade-off, i.e., when power is concentrated on fewer beams to maximize sensing accuracy (i.e., lower miss detection), the signal's entropy is reduced, thereby limiting its communication capacity. Conversely, distributing power across multiple beams increases communication diversity but leads to noisier sensing decisions. 
For $\Qsf >1$, the performance varies based on system parameters (e.g., power constraint, signal dimension, blocklength, and feedback power threshold), with a small gap observed between the inner and outer bounds. While fixing the number of beams selected at each transmission, as specified in~\eqref{eq:semi-bi-sec-d-update}, reduces the complexity of the optimization problem in Proposition~\ref{lemma:bi-section}, it results in a degradation of sensing performance.  

     \section{Conclusion}
     
     In this work, we analyzed Gaussian beam-pointing channels with block memory and binary feedback, characterizing the key aspects of the beam acquisition problem.  We focused on maximizing sensing performance under per-slot power constraints.  By formulating the sensing capacity as a finite-horizon dynamic programming problem, we proposed a coded beam acquisition scheme that dynamically updates transmission policies based on posterior state probabilities. This approach enables the joint use of Gaussian codewords for both sensing and communication, demonstrating the effectiveness of feedback-driven beam selection, particularly under short coherence blocklengths. 
      The special case of memoryless channels highlighted the fundamental trade-off between sensing accuracy and communication rate, captured by the mutual information term   $I(\underline S; Z^{\Qsf} | \underline X^{\Qsf})$, which reflects the precision of the posterior state estimate. In JCAS design, such metrics also influence the communication codebook design since probing affects the signal's structure. Therefore, our formulation bridges pure state sensing models with broader JCAS frameworks, offering a flexible approach to performance optimization depending on system priorities.

\clearpage

\bibliographystyle{IEEEtran}
\bibliography{refs-isac-v2}
\end{document}

%% file: macros.tex
\newfont{\bbb}{msbm10 scaled 700}

\newfont{\bb}{msbm10 scaled 1100}

\newcommand{\RR}{\mbox{\bb R}}

\newcommand{\EE}{\mbox{\bb E}}

\newcommand{\Bc}{{\cal B}}

\newcommand{\Rc}{{\cal R}}
\newcommand{\Sc}{{\cal S}}

\usepackage{stmaryrd} 

\newcommand{\BLUE}{\color[rgb]{0,0,0.90}}

\usepackage{hyperref}
\hypersetup{
    bookmarks=true,         
    unicode=false,          
    pdftoolbar=true,        
    pdfmenubar=true,        
    pdffitwindow=false,     
    pdfstartview={FitH},    
    pdfnewwindow=true,      
    colorlinks=true,       
    linkcolor=red,          
    citecolor=green,        
    filecolor=blue,      
    urlcolor=blue           
}

\newcommand{\mk}{{\rm -\!o\!-}}

%% file: authors.tex
\author{\IEEEauthorblockN{Siyao Li\IEEEauthorrefmark{1}\IEEEauthorrefmark{2}, 
Shuangyang Li\IEEEauthorrefmark{2}, 
		 Giuseppe Caire\IEEEauthorrefmark{2}}
	\IEEEauthorblockA{
	\IEEEauthorrefmark{1}Embry-Riddle Aeronautical University, Daytona Beach, USA\\
		\IEEEauthorrefmark{2}Technical University of Berlin, Germany\\
		Emails: lis14@erau.edu , \{shuangyang.li, caire\}@tu-berlin.de}}